\documentclass[aps,prl,amsmath,amssymb,twocolumn]{revtex4}
\usepackage{amsmath}
\usepackage{amsfonts}
\usepackage{ulem}
\usepackage{color}
\usepackage{float}

\usepackage{soul}
\usepackage{graphicx}
\usepackage{amssymb}
\usepackage{epstopdf}
\usepackage{url}
\usepackage{setspace}

\newcommand{\be}{\begin{equation}}
\newcommand{\ee}{\end{equation}}
\newcommand{\ber}{\begin{eqnarray}}
\newcommand{\eer}{\end{eqnarray}}

\begin{document}
	\title{Electrically Tuneable Nonequilibrium Optical Response of Graphene}
	
	\author{Eva A. A. Pogna$^{1,2,*}$, Andrea Tomadin$^{3}$, Osman Balci$^{4}$, Giancarlo Soavi$^{4,5,6}$, Ioannis Paradisanos$^{4}$, Michele Guizzardi$^1$, Paolo Pedrinazzi$^{7}$, Sandro Mignuzzi$^{4}$, Klaas-Jan Tielrooij$^{8}$, Marco Polini$^{3}$, Andrea C. Ferrari$^{4}$, Giulio Cerullo$^{1,9}$}

	\affiliation{$^{1}$ Dipartimento di Fisica, Politecnico di Milano, P.zza Leonardo da Vinci 32, Milano 20133, Italy\\
		$^{2}$ NEST, Istituto Nanoscienze-CNR and Scuola Normale Superiore, P.zza S. Silvestro 12, Pisa 56127, Italy\\
		$^{3}$ Dipartimento di Fisica, Universit{\`a} di Pisa, Largo Bruno Pontecorvo 3,  Pisa 56127, Italy\\
		$^{4}$ Cambridge Graphene Centre, University of Cambridge, 9 JJ Thomson Avenue, Cambridge CB3 0FA, UK\\
		$^{5}$ Institute of Solid State Physics, Friedrich Schiller University Jena, Jena 07743, Germany\\
		$^{6}$ Abbe Center of Photonics, Friedrich Schiller University Jena, Jena 07745, Germany\\
		$^7$ L-NESS, Department of Physics, Politecnico di Milano, Via Anzani 42, Como 22100, Italy\\
		$^8$ Catalan Institute of Nanoscience and Nanotechnology (ICN2), BIST \& CSIC, Campus UAB, Bellaterra (Barcelona) 08193, Spain\\
		$^9$ Istituto di Fotonica e Nanotecnologie, Consiglio Nazionale delle Ricerche, P.zza L. da Vinci 32, Milano 20133, Italy}


\begin{abstract}
The ability to tune the optical response of a material \textit{via} electrostatic gating is crucial for optoelectronic applications, such as electro-optic modulators, saturable absorbers, optical limiters, photodetectors and transparent electrodes. The band structure of single layer graphene (SLG), with zero-gap, linearly dispersive conduction and valence bands, enables an easy control of the Fermi energy $E_{\rm F}$ and of the threshold for interband optical absorption. Here, we report the tunability of the SLG non-equilibrium optical response in the near-infrared (1000-1700nm/0.729-1.240eV), exploring a range of $E_{\rm F}$ from -650 to 250meV by ionic liquid gating. As $E_{\rm F}$ increases from the Dirac point to the threshold for Pauli blocking of interband absorption, we observe a slow-down of the photobleaching relaxation dynamics, which we attribute to the quenching of optical phonon emission from photoexcited charge carriers. For $E_{\rm F}$ exceeding the Pauli blocking threshold, photobleaching eventually turns into photoinduced absorption, due to hot electrons' excitation increasing SLG absorption. The ability to control both recovery time and sign of non-equilibrium optical response by electrostatic gating makes SLG ideal for tunable saturable absorbers with controlled dynamics.
\end{abstract}
\maketitle

Single layer graphene (SLG) has peculiar optoelectronic properties\cite{ferrari2015science, bonaccorso2010graphene,romagnoli2018graphene}, which stem from the physics of its massless Dirac fermions. These include high electron mobility ($>$100,000 cm$^{2}$V$^{-1}$s$^{-1}$ at room temperature (RT)\cite{Du2008,Bolotin2008,purdie2018cleaning,defazio2019high}), broadband optical absorption\cite{Nair2008}, tunability of Fermi energy $E_{\rm F}$ \textit{via} electrostatic gating\cite{pisana2007} resulting from the linear dispersion of conduction (CB) and valence bands (VB), and a vanishing density of states at the Dirac point\cite{Neto2009}.

Light absorption in SLG is due to the interplay of intraband\cite{Ando2002,Gusynin2006,Horng2011} and interband\cite{Mak2008,Mak2012} transitions. In undoped SLG, the first ones dominate in THz\cite{Liu2014} and microwaves\cite{Balci2017}, the second\cite{Mak2012} in near-infrared (NIR)\cite{Mak2008} and visible (VIS)\cite{Mak2012}. Electrical control of $E_{\rm F}$, by exploiting the band-filling effect\cite{Fengwang2008}, allows one to vary the density of electronic states available for both intraband\cite{Mak2008} and interband transitions\cite{Fengwang2008,romagnoli2018graphene}, thus affecting the linear absorption of SLG over a broad range from THz\cite{HighSTHz2015,HighSTHz2014,SiGrTHz2014,Kakenov2016,DiGaspare2020} to NIR\cite{Dalir2016,CapassoIR2014,ChenIR2018,Romagnoli2019} and VIS\cite{Coskun2013}. This led to the development of SLG-based electro-optic modulators\cite{FengWang2011,Sorianello2018,romagnoli2018graphene,CapEOM2014,liuEOM2014,Romagnoli2019,Dalir2016,CapassoIR2014,LipsonIR2015}, which can reach higher modulation speed (up to 200GHz\cite{Liu2020}) than LiNbO$_3$\cite{Li2020} and Si\cite{Rahim2021}, due to the superior mobility of SLG charge carriers, with high modulation depths both in amplitude (up to$\sim60\%$)\cite{HighSTHz2015,SiGrTHz2014,CapassoIR2014,ChenIR2018,romagnoli2018graphene} and phase ($\sim 65^\circ$)\cite{Sorianello2018,romagnoli2018graphene}.

SLG also exhibits large nonlinear optical response\cite{Jiang2018,Soavi2018,yoshikawa2017,hafez2018,Kovalev2021}, due to a strong coupling to light. The third-order nonlinear optical susceptibility of SLG in the NIR at 0.7eV is\cite{Soavi2018} $\chi^3\sim$5$\times$10$^{-18}$ m$^2$V$^{-2}$, several orders of magnitude higher than in dielectrics (\textit{e.g.} $\sim10^{-22}$m$^2$V$^{-2}$ for SiO$_2$\cite{Buchalter1982}) and atomically thin semiconductors (\textit{e.g.} $\sim6\times10^{-20}$m$^2$V$^{-2}$ for single layer WSe$_2$\cite{Rosa2018}). Nonlinearities of order higher have been exploited for high-harmonics generation in SLG\cite{yoshikawa2017,hafez2018}. The strong nonlinear response results also in saturable absorption\cite{Sun2010}, optical Kerr effect\cite{Yu2016}, and optical bistability\cite{Peres2014,Sadeghi2018}, \textit{i.e.} the ability to provide two stable optical outputs for a specific light input\cite{Abraham1982}.  $E_{\rm F}$ control \textit{via} external gating allows one to tune the nonlinear optical response of SLG, resulting in gate-tunable third-harmonic generation\cite{Jiang2018,Soavi2018,Kovalev2021} and four-wave-mixing\cite{Alexander2017}.

The $E_{\rm F}$ dependence of the transient absorption properties of SLG when brought out of equilibrium remains still largely unexplored, with studies limited to the THz range\cite{Frenzel2014,Shi2014,Hafez2015,Lin2015}, discussing the tuning of intraband photoconductivity with $E_{\rm F}$\cite{Frenzel2014,Shi2014,Hafez2015}. The modulation of interband absorption in NIR and VIS is more challenging to study due to the need of $E_{\rm F}\sim$0.5eV in order to cross the Pauli blocking threshold, above which the non-equilibrium optical properties have been only theoretically explored\cite{mikhailov2019theory}.

The non-equilibrium optical response of SLG is crucial for optoelectronic applications, such as photodetection\cite{Koppens2014}, relying on the relaxation dynamics of photoexcited charge carriers. Numerous ultrafast optical spectroscopy experiments were performed on SLG\cite{dawlaty2008,sun2008,Breusing2011,Kotov2012,Brida2013} to investigate the charge-carriers relaxation dynamics by looking at the modifications it induces on SLG absorption. In a pump-probe experiment, the system is photoexcited by an optical pulse, the pump, whose duration is to be shorter than the timescale of the relaxation processes under investigation. The relaxation of the photoexcited system is then monitored by detecting the absorption of a second optical pulse, the probe, as a function of the time delay with respect to the pump pulse\cite{Cerullo2007}.

In SLG, interband absorption of the pump pulse induces out-of-equilibrium distributions of holes (h) and electrons (e) in VB and CB, respectively, peaked at $\pm\hbar\omega_{pump}/2$, where $\hbar\omega_{pump}$ is the pump photon energy. Carrier-carrier scattering drives the ultrafast e-h thermalization on a time-scale $\tau_{th}<20$fs\cite{Brida2013} from out of equilibrium, to an hot Fermi-Dirac distributions (HFD) with defined electronic temperature $T_{\rm e}$. The HFD can be detected in a pump-probe experiment as a photobleaching (PB) signal\cite{dawlaty2008,sun2008,Breusing2011}, \textit{i.e.} decreased probe absorption compared to equilibrium, due to Pauli blocking of interband transitions caused by the photo-generated e/h. The excess energy of the hot charge-carriers is released to the lattice \textit{via} electron-phonon scattering with optical phonons\cite{Lazzeri2005,Tomadin2013,Pogna2021}, anharmonically coupled to acoustic phonons\cite{Lazzeri2005,Bonini2007,Tomadin2013,Pogna2021}. Hot carriers cooling occurs on a few-ps time-scale\cite{Bonini2007,dawlaty2008,sun2008,Breusing2011,Brida2013} and is influenced, through the activation of additional relaxation channels, by the dielectric environment (\textit{e.g.} \textit{via} near-field coupling to hyperbolic optical phonons of the substrate or encapsulant material\cite{tielrooij2018}). Defects can also accelerate the cooling \textit{via} electron-phonon interaction by acting as scattering centres mediating the direct coupling of the hot charge carriers with finite momentum acoustic phonons\cite{graham2013photocurrent,betz2013supercollision,Song2012}. This process, referred to as  supercollision\cite{graham2013photocurrent,betz2013supercollision,Song2012}, accelerates the cooling for increasing defect density\cite{alencar2014}.

Here we investigate the $E_{\rm F}$ dependence of the non-equilibrium optical response of SLG in the NIR range between 0.729 and 1.240eV (1000-1700nm), exploiting ionic liquid gating to tune $E_{\rm F}$ from -650 to 250meV, thus exceeding the Pauli blocking threshold for interband absorption, achieved when $|E_{\rm F}|=\hbar\omega_{probe}/2$, where $\hbar\omega_{probe}$ is the energy of the probe beam. Applying ultrafast pump-probe spectroscopy with 100fs time resolution, we detect the changes with $E_{\rm F}$ of amplitude and sign of the differential transmission ($\Delta T/T$), as well as of its relaxation dynamics. Starting from not intentionally doped SLG and increasing $E_{\rm F}$, we first observe a rise in PB amplitude ($\Delta T/T >$0) with a slow-down of its relaxation dynamics. Above the Pauli blocking  threshold, photoexcitation has an opposite effect on SLG, activating additional absorption channels, as shown by the appearance of photoinduced absorption (PA) ($\Delta T/T<$0). The $\Delta T/T$ changes are assigned to the $E_{\rm F}$ dependence of the hot carriers cooling dynamics, simulated considering relaxation through the emission of optical phonons. The gate tunability of the non-equilibrium optical response is key for optoelectronic applications, such as saturable absorbers (SA) with gate-tunable response.

\begin{figure}
	\centerline{\includegraphics[width=80mm]{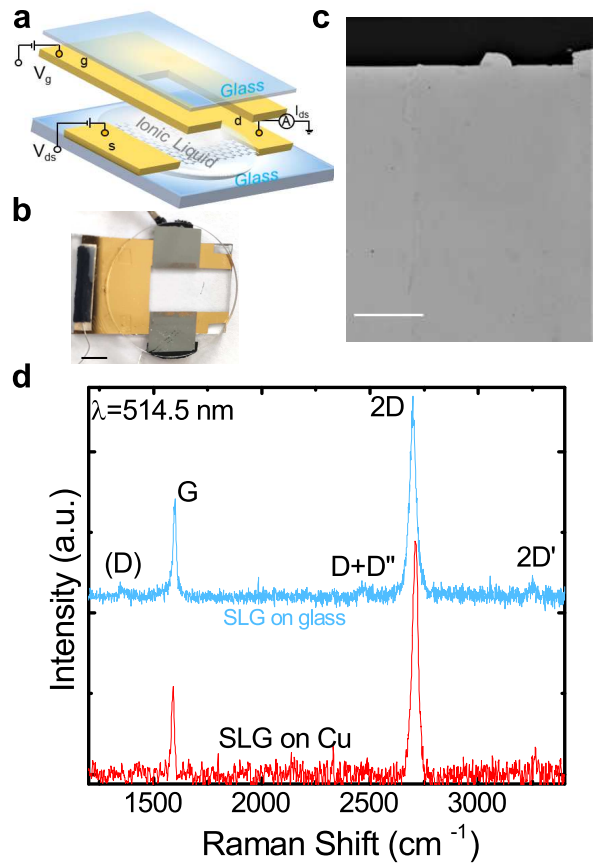}}
	\caption{(\textbf{a}) Schematic of device with source (s), drain (d) and gate (g) contacts used to tune the SLG $E_{\rm F}$ while measuring its transmission properties. (\textbf{b}) Photo of representative device. Scale bar 5mm. (\textbf{c}) Image of transferred SLG in a region near to the drain contact (dark area on the top). Scale bar: 100 $\mu$m. (\textbf{d}) 514.5 nm Raman spectrum of SLG as-grown and transferred on glass.}
	\label{f:gatingT1}
\end{figure}

\section{Results and Discussion}
We modulate $E_{\rm F}$ by means of the electrostatic field effect\cite{Dop2008} using an ionic-liquid top-gated field effect transistor (FET) sketched in Fig.\ref{f:gatingT1}a. The top-gate geometry, with Diethylmethyl (2-methoxyethyl) ammoniumbis-(trifluoromethylsulfonyl)imide ($C_6H_{20}F_6N_{2}O_5S_2$) as ionic liquid, is chosen to allow light measurements in transmission through an$\sim$1cm$^2$ optical window.  Large area (8mm$\times$8mm) SLG is prepared by chemical vapor deposition (CVD) as for Ref.\citenum{Li2009}. The device fabrication follows  Ref.\citenum{soavi2019hot}. Fig.\ref{f:gatingT1}b is a photo of the device, and Fig.\ref{f:gatingT1}c an optical image of the transferred SLG, showing no macroscopic tearing nor folding.

Both as-grown and transferred SLG are characterized with a Renishaw InVia Raman spectrometer using a 50x objective, a CW laser at 514.5nm, with power on the sample$<$0.5mW to exclude heating effects. The Raman peaks are fitted with Lorentzians, with error bars derived from the standard deviation across 6 spectra and the spectrometer resolution $\sim$1cm$^{-1}$. The Raman spectrum of as-grown SLG on Cu is in Fig.\ref{f:gatingT1}d, after Cu photoluminescence removal\cite{CuPL2013}. The 2D peak is a single Lorentzian with full-width half maximum FWHM(2D)$\sim$31$\pm$3cm$^{-1}$, signature of SLG\cite{SLG2006}. The G peak position Pos(G) is$\sim$1586$\pm$2cm$^{-1}$, with FWHM(G)$\sim$16$\pm$3cm$^{-1}$. The 2D peak position, Pos(2D), is$\sim$2704$\pm$4cm$^{-1}$, while the 2D to G peak intensity and area ratios, I(2D)/I(G) and A(2D)/A(G), are 3.1$\pm$0.4 and 6.2$\pm$0.7. No D peak is observed, indicating negligible Raman active defects\cite{Ferrari2000,Ferrari2013}.

The Raman spectrum of SLG transferred on glass is in Fig.1d. The 2D peak retains its single-Lorentzian line shape with FWHM(2D)$\sim$36$\pm$1cm$^{-1}$. Pos(G)$\sim$1597$\pm$1cm$^{-1}$, FWHM(G) $\sim$15$\pm$1cm$^{-1}$, Pos(2D)$\sim$2696$\pm$1cm$^{-1}$, I(2D)/I(G)$\sim$2$\pm$0.2 and A(2D)/A(G)$\sim$4.9$\pm$0.3 indicating \textit{p-}doping with $E_{\rm F}\sim$-230$\pm$80meV\cite{Dop2008,EEDop2009}. I(D)/I(G)$\sim$0.06$\pm$0.05 corresponds\cite{Bruna2014} to a defect density $\sim$2.6$\pm$1.9$\times$10$^{10}$cm$^{-2}$ for excitation energy 2.410eV (514.5nm) and $E_{\rm F}$=-230$\pm$80meV. Pos(G) and Pos(2D) are affected by the presence of strain\cite{Mohiuddin2009}. For uniaxial(biaxial) strain, Pos(G) shifts by $\Delta$Pos(G)/$\Delta\varepsilon\sim$23(60)cm$^{-1}$\%$^{-1}$\cite{Mohiuddin2009,Yoon2011}. Pos(G) also depends on $E_{\rm F}$\cite{pisana2007,Dop2008}. The average doping as derived from A(2D)/A(G), FWHM(G) and I(2D)/I(G), should correspond to Pos(G)$\sim$1588$\pm$1cm$^{-1}$ for unstrained graphene\cite{pisana2007,Dop2008}. However, we have Pos(G)$\sim$1597$\pm$1cm$^{-1}$, which implies a contribution from uniaxial (biaxial) strain$\sim$0.16$\pm$0.02\% (0.4$\pm$0.04\%)\cite{Mohiuddin2009,Yoon2011}.

\begin{figure}
\centerline{\includegraphics[width=78mm]{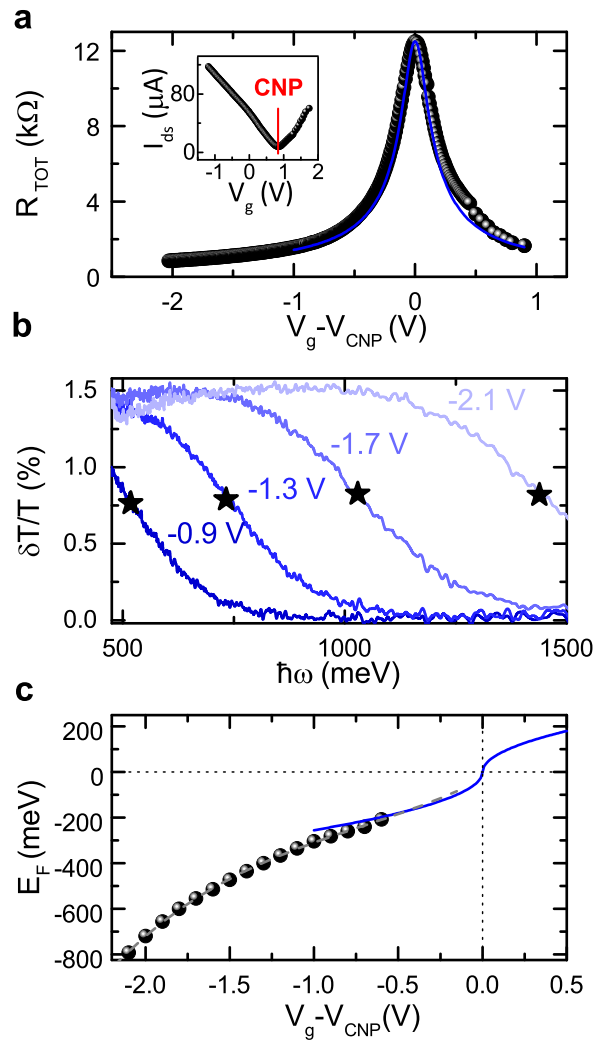}}
\caption{(\textbf{a}) Total resistance $R_{\rm TOT}$ as a function of $V_{\rm g}$-$V_{\rm CNP}$ (black dots) with Drude model fitting (blue solid line) to estimate the residual charge carrier density $n_0$ and gate capacitance $C$. Inset, $I_{\rm ds}$ for $V_{\rm ds}$= 100mV. (\textbf{b}) $\delta T/T$ for different $V_g$-$V_{\rm CNP}$ (indicated next to the curves) as a function of photon energy $\hbar\omega$, showing the gate tunability of the absorption edge for interband transitions; (\textbf{c}) $E_{\rm F}$ determined from $\delta T/T$ as a function of $V_g$-$V_{\rm CNP}$ (full dots) interpolated by f(x)= $A_0$+$A_1$x+$A_2$x$^{2}$+$A_3$x$^{3}$ with $A_0$=-0.03eV, $A_1$=0.37eV/V $A_2$=0.17eV/V$^{2}$ $A_3$=0.08eV/V$^{3}$ (grey dashed line), together with the trend (blue solid line) $E_{\rm F}=\hbar v_{\rm F}\sqrt{\pi n_e(V_{\rm g})}$ where $n_e(V_{\rm g})= (V_{\rm g}-V_{\rm CNP})(C/e)$ is the gate-tuneable charge carrier density}
\label{f:gatingT2}
\end{figure}

The gate voltage $V_{\rm g}$ polarizes the ionic liquid leading to the formation of electrical double layers (EDLs), near the SLG and
Au interfaces\cite{Dop2008,DasPRB}, that modulate the carrier density. Since the EDL thickness is$\sim$1nm for ionic liquids\cite{Ye2010,EDLTs2015}, the solid-liquid interfacial electric field and the induced charge densities on the surface reach values\cite{Ye2010,Kakenov2016} as large as$\sim$10-20MVcm$^{-1}$ and $10^{14}$cm$^{-2}$ even at moderate $V_{\rm g}\sim$1-2V. The transfer characteristics of our device for source-drain bias $V_{\rm ds}$=100mV is in Fig.\ref{f:gatingT2}a. This exhibits a typical ambipolar behaviour, as seen by the \textit{V}-shaped gate dependence of the source-drain current $I_{\rm ds}$. The channel resistance peaks at $V_{\rm CNP}$=0.84V, corresponding to the charge neutrality point (CNP), where the density of states in SLG reaches its minimum\cite{Geim2007,Kakenov2016,DiGaspare2020}. $V_{\rm CNP}$ depends on $E_{\rm F}$, on the gate-metal work function\cite{Kakenov2016} and on the choice of contact materials\cite{Xu2011}.

In order to determine $E_{\rm F}$ as a function of $V_{\rm g}$, we measure the static transmission $T$ in the NIR (500-1500meV) with an Agilent Cary 7000 UV-VIS-MIR spectrometer. Fig.\ref{f:gatingT2}b plots a selection of transmission spectra for different $V_g$, compared to that at the CNP, evaluated as follows: $\delta T/T=\frac{T(V_{\rm g})-T(V_{\rm CNP})}{T(V_{\rm CNP})}$. $T$ increases with respect to the CNP, \textit{i.e.} $\delta T/T>0$, when absorption is inhibited by Pauli blocking, due to e in CB (\textit{n-}doping) or h in VB (\textit{p-}doping). In terms of probe photon energy, this corresponds to $\hbar\omega_{probe}<2|E_{\rm F}|$. We estimate $E_{\rm F}$ considering that $\delta T/T$ halves\cite{soavi2019hot} for $\hbar\omega_{probe}=2|E_{\rm F}|$ at values indicated by black stars in Fig.\ref{f:gatingT2}b. For probe photon energies $\hbar\omega_{probe}<2|E_{\rm F}|$, interband absorption is blocked and the sample has $T\sim$99.6-99.8$\%$, with$\sim$0.2-0.4$\%$ residual absorption attributed to intraband transitions enabled by disorder\cite{soavi2019hot}. The $T$ modulation due to the bleaching of interband absorption is $\sim$1.5$\%$ against the $\sim$2.3$\%$ expected for suspended SLG\cite{Nair2008}, because of the presence of the glass substrate and of the Diethylmethyl (2-methoxyethyl) ammoniumbis-(trifluoromethylsulfonyl)imide\cite{Lidorikis}, with refractive index$\sim$1.418-1.420\cite{aldrich}.

$E_{\rm F}$ extracted from the $T$ measurements is plotted in Fig.\ref{f:gatingT2}c as a function of $V_{\rm g}$. At $V_{\rm g}$=0V there is a \textit{p-}doping $E_{\rm F}\sim$-250meV in agreement with the Raman estimation ($\sim$-230$\pm$80meV) without ionic liquid. From the analysis of the charge-transfer curve with the Drude model as in Ref.\citenum{DiGaspare2020} (see blue line in Fig.\ref{f:gatingT2}a), we evaluate a residual charge carrier density $n_0$=5.6$\times$10$^{11}$cm$^{-2}$, responsible for the finite conductivity at the CNP, and a gate capacitance $C$= 766nFcm$^{-2}$. We note this is a typical $n_0$ for as grown and transferred SLG\cite{Casiraghi2018}. Lower residual doping $\sim10^{11}cm^{-2}$ can be achieved with cleaning techniques\cite{purdie2018cleaning}, not used here. The finite electrical conductivity and doping at the CNP\cite{Kakenov2016,DiGaspare2020} are due to electron-hole puddles\cite{Martin2007} at the micrometer scale, caused by charged impurities\cite{Adam2007} located either in the dielectric, or at the SLG/dielectric interface\cite{Adam2007}.
Near the CNP, for $|V_{\rm g}$-$V_{\rm CNP}|<$0.6V, the interband absorption edge is outside the spectral window of our $\delta T/T$ measurements and we evaluate\cite{Nair2008} $E_{\rm F}=\hbar v_{\rm F}\sqrt{\pi n_e(V_{\rm g})}$, with $v_{\rm F}$ the Fermi velocity, directly from the gate tunable charge carrier density $n_e(V_{\rm g})= (V_{\rm g}-V_{\rm CNP})(C/e)$, with $e$ the electron charge, see blue solid line in Fig.\ref{f:gatingT2}c. At high gate voltages $|V_{\rm g}-V_{\rm CNP}|>$ 0.9V, the disagreement between the calculated $E_{\rm F}$ and the values obtained from $\delta T/T$, is attributed to the dependence of mobility on charge carrier density\cite{purdie2018cleaning}, not included in the analysis of the transport properties at the base of the calculated $E_{\rm F}$, that is valid near the CNP and fails to describe the sample behavior for $|V_{\rm g}-V_{\rm CNP}|>$ 0.9V. Accordingly, we assume the values extracted from $\delta T/T$ (Fig.\ref{f:gatingT2}c grey dashed line) in the range $|V_{\rm g}-V_{\rm CNP}|>$ 0.6V, and those calculated from $n_e(V_{\rm g})$ (Fig.\ref{f:gatingT2}c, blue solid line), in the range $|V_{\rm g}-V_{\rm CNP}|<$ 0.6V.
At each $V_{\rm g}$ of Fig.\ref{f:gatingT2}, we also monitor the source-drain current $I_{\rm ds}$ (inset of Fig.\ref{f:gatingT2}a) to determine the empirical relation with $E_{\rm F}$. The transfer curve that we obtain allows us to track $E_{\rm F}$ by monitoring $I_{\rm ds}$ during all the subsequent measurements. We test the $E_{\rm F}$ tunability of the ionic liquid top-gate device up to -800meV, corresponding to a wide range of charge carrier densities from $\sim4.5\times10^{12}$ cm$^{-2}$ (\textit{n-}doping) to $\sim -4.7\times10^{13}$ cm$^{-2}$ (\textit{p-}doping), much wider than possible with a 285nm SiO$_2$ back gate, usually limited to $\pm6\times10^{12}$cm$^{-2}$ by the gate capacitance\cite{Geim2007}. We got similar $E_{\rm F}(V_g)$ in Ref.\citenum{soavi2019hot} from Raman spectra and NIR transmission.
\begin{figure*}
\centerline{\includegraphics[width=150mm]{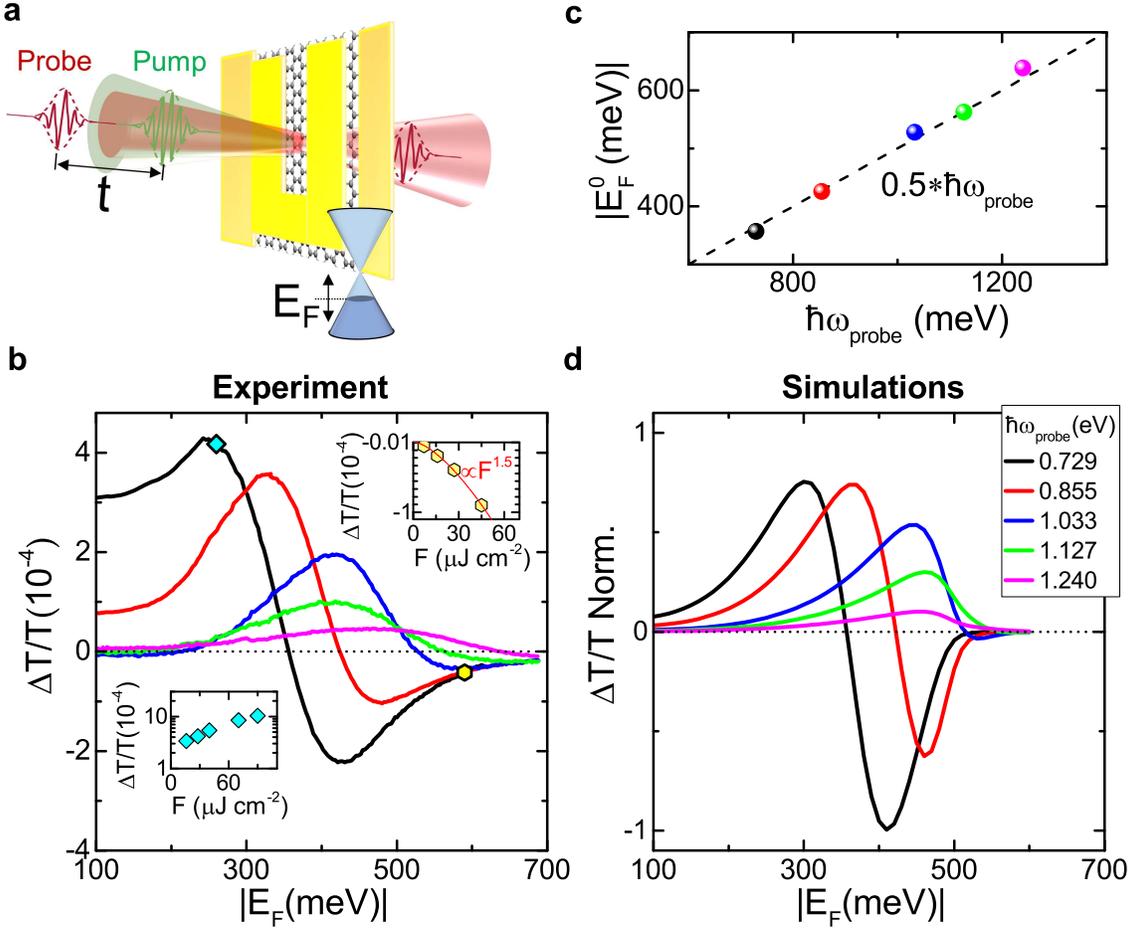}}
\caption{(\textbf{a}) Sketch of pump-probe experiment on SLG with tunable $E_{\rm F}$ controlled by V$_{\rm g}$. \textbf{(b)} Experimental $\Delta T/T$ at $t$= 150fs as a function of $|E_{\rm F}|$ for \textit{p-}doping acquired at different $\hbar\omega_{probe}$=0.729, 0.855, 1.033, 1.127, 1.240eV (legend of panel d) for a pump fluence F$\sim$28$\mu$J cm$^{-2}$. Top-right inset: fluence dependence of $\Delta T/T$ amplitude above Pauli blocking for pump absorption, at $\hbar\omega_{probe}$=0.729eV and $E_{\rm F}$=-590meV (hexagonal yellow symbol in main figure),  together with a superlinear power-law dependence on F (solid red line). Bottom-left inset: fluence dependence of $\Delta T/T$ amplitude, for $\hbar\omega_{probe}$=0.729eV and $E_{\rm F}$=-260meV (rhombus cyan symbol in main figure). \textbf{(c)} Fermi energy $|E_{\rm F}^0|$ at $\Delta T/T=0$, extracted from panel b, as a function of $\hbar\omega_{probe}$. \textbf{(d)} Simulated $\Delta T/T$ at t=150fs as a function of $|E_{\rm F}|$ for \textit{p-}doping at the same $\hbar\omega_{probe}$ as in panel b.}
\label{f:gatingDT}
\end{figure*}

We perform ultrafast pump-probe spectroscopy as sketched in Fig.\ref{f:gatingDT}a. The pump is a 100fs NIR pulse centred at $\hbar\omega_{pump}=$0.8eV, while the probe spectrum covers $\hbar\omega_{probe}$=0.729-1.240eV (see Methods for details). The relaxation dynamics is monitored through the differential transmission $\Delta T(t)/T=\frac{T_{\rm pump-ON}(t)-T_{\rm pump-OFF}}{T_{\rm pump-OFF}}$ evaluated from the probe transmission with ($T_{\rm pump-ON}$) and without ($T_{\rm pump-OFF}$) pump excitation, after a time delay $t$ between probe and pump pulses, varied with an optical delay line. Given that the pulses duration exceeds the time-scale of carrier-carrier thermalization\cite{Brida2013}, we can assume charge carriers thermalized to HFDs, and investigate their cooling dynamics.

Fig.\ref{f:gatingDT}b plots $\Delta T/T$ at $t$=150fs, chosen as the delay at which the maximum signal amplitude is reached for $V_{\rm g}$=0V. The signal is plotted as a function of $E_{\rm F}$ for different probe photon energies. Since the transient response is symmetric with respect to the CNP for \textit{n-} and \textit{p-} doping and our SLG is \textit{p-}doped at $V_{\rm g}$=0, we explore negative $E_{\rm F}$ in order to reach higher $|E_{\rm F}|$ by applying a smaller $V_{\rm g}$. We observe a strong modulation of $\Delta T/T$ with $E_{\rm F}$, higher at the low energy tail of the probe pulse, with the signal changing from 4 to -2$\times 10^{-4}$ (see the curve at 0.729eV in Fig.\ref{f:gatingDT}b). The signal amplitude decreases for increasing $\hbar\omega_{probe}$, as expected for a thermal distribution of carriers\cite{Breusing2011}. In all the probed range, near the CNP, we observe, as expected, a PB signal, \textit{i.e.} $\Delta T/T(t)>0$.  By increasing $|E_{\rm F}|$, first PB increases in amplitude, then a change of sign occurs at a threshold $|E_{\rm F}|$ dependent on the probe photon energy. The Fermi energy at which the sign change occurs, $|E_{\rm F}^0|$ in Fig.\ref{f:gatingDT}c, corresponds to $\hbar\omega_{probe}/2$, \textit{i.e.} the Pauli blocking threshold for probe photons. Above this, the pump pulse, exciting e (h) to higher (lower) energy states, partially unblocks the probe interband absorption, otherwise inhibited, resulting in a PA signal, \textit{i.e.} $\Delta T/T<0$. The PA intensity increases with $E_{\rm F}$ up to a peak, whose position in terms of $E_{\rm F}$ increases with probe photon energy. A constant $\Delta T/T\sim-1\times10^{-5}$ is then approached in the high $|E_{\rm F}|$ limit ($E_{\rm F}<$-690meV) in all the probed range.

We note that Ref.\citenum{Katayama2020} reported a study of $E_{\rm F}$ dependence of the transient optical properties of SLG, without reaching Pauli blocking, \textit{i.e.} $|E_{\rm F}|>\hbar\omega_{probe}/2$, which we investigate here, revealing the PA regime. Our results show that, by varying $V_{\rm g}$, we can not only control the relaxation dynamics of SLG, but also change the $\Delta T/T$ sign.

Above the Pauli blocking threshold for pump interband transitions ($|E_{\rm F}|\geq$400meV for $\hbar\omega_{pump}$=800meV), $\Delta T/T$ is expected to vanish, because the pump should not be able to photoexcite SLG. However, a finite value is observed, caused by residual pump absorption, related to both extrinsic\cite{Basov2008,Mak2012,soavi2019hot} and intrinsic\cite{Mak2012,Kittel} effects. Amongst the former, charged impurities and scatterers (\textit{e.g.} edge defects, cracks, vacancies) can induce residual conductivity\cite{Basov2008,soavi2019hot} activating intraband absorption. Amongst the latter, is the residual absorption from the tail of the carrier Fermi distribution, \textit{i.e.} off-resonance absorption, which has a finite broadening at RT\cite{Kittel}. The fluence dependence of $\Delta T/T$ at $\hbar\omega_{probe}=$0.729eV in the inset of Fig.\ref{f:gatingDT}b, is superlinear above the threshold for Pauli blocking of pump absorption (as measured at $|E_{\rm F}|=$590meV), suggesting a non-negligible contribution from two-photon absorption\cite{marini2017}. This could also explain the vanishing signal when approaching $|E_{\rm F}|=$800meV (the Pauli blocking threshold for two-photon absorption). 

While height and width of PB and PA bands slightly change with $\hbar\omega_{probe}$, we observe similar features in all the probed range upon increasing $|E_{\rm F}|$: an increase of PB, followed by a decrease, and a sign change above the Pauli blocking threshold for probe absorption. The measurements are performed using a low excitation fluence (28$\mu$Jcm$^{-2}$) to work in a perturbative regime corresponding to $T_{\rm e}<$1000K, thus reducing the impact of the hot-phonon bottleneck and focusing on the electron-optical phonon cooling. The amplitude of the $\Delta T/T$ signal increases by increasing excitation fluence, as shown in the inset of Fig.\ref{f:gatingDT}b, by almost two orders of magnitude for the PA signal (see top-right inset of Fig.3b at  $|E_{\rm F}|$=590meV) and one order of magnitude for the PB signal (see bottom-left inset of Fig.3b at $|E_{\rm F}|$=260meV). We previously reported the use of ungated SLG as SA (GSA) in mode-locked lasers in which an absorption modulation$\leq 1.3\%$ is sufficient to induce and control the pulsed (mode-locked) regime\cite{Sun2010}. The ability to electrically control amplitude, sign and recovery time of $\Delta T/T$ in a GSA is thus of immediate practical relevance for optimizing mode-locking regime for stability, pulse width and average output power.
\begin{figure*}[ht]
	\centerline{\includegraphics[width=150mm]{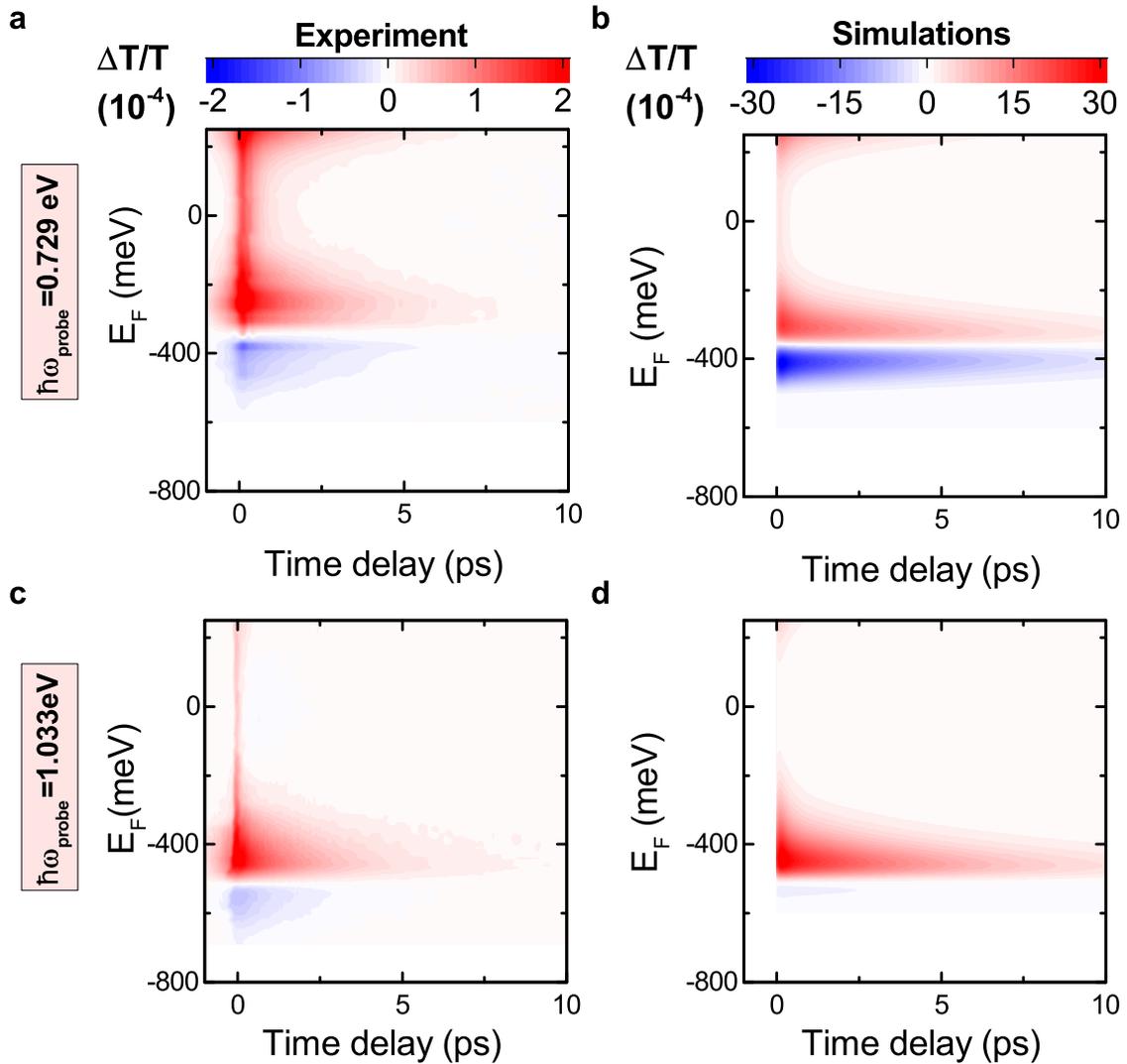}}
	\caption{Time-evolution of $\Delta T/T (t)$ for different $E_{\rm F}$ at (\textbf{a},\textbf{b}) $\hbar \omega_{probe}$=0.729eV, with (a) experiment, and (b) simulations, and (\textbf{c},\textbf{d}) $\hbar \omega_{probe}$=1.033eV, with (c) experiment and (d) simulations. $t=0$ corresponds to the pump arrival. The signal at negative delays indicates a finite build-up time, exceeding the pump-probe time duration at $|E_{\rm F}|$=250meV for $\hbar\omega_{probe}$=0.729eV and 350meV for $\hbar\omega_{probe}$=1.033eV, approaching the Pauli blocking $E_{\rm F}$.}
	\label{f:dyn}
\end{figure*}

To understand the $E_{\rm F}$ dependence of the non-equilibrium optical response of SLG, we calculate $\Delta T/T$ (see Methods for details) as a function of initial carrier density $n_e$, related\cite{Nair2008} to $E_{\rm F}$ by $n_{\rm e}=\frac{1}{\pi}(\frac{E_{\rm F}}{\hbar v_{\rm F}})^2$. $\Delta T/T$ in Fig.\ref{f:gatingDT}d is computed from the changes in optical conductivity $\Delta\sigma$ induced by photoexcitation as a function of $E_{\rm F}$. To evaluate $\Delta T/T$ at $t$=150fs we consider the charge carriers as distributed in energy and momentum along a HFD with a time-dependent chemical potential $\mu_c$ and $T_e(t)>$RT. Our model takes into account that, even though the pump fluence is constant, the initial $T_{\rm e}$ changes with $E_{\rm F}$ due to the change of pump absorption. We consider the absorption from the tail of the Fermi-Dirac distribution as source of residual pump absorption for $|E_{\rm F}|>$400meV. The charge carrier distribution modification with $E_{\rm F}$ is sufficient to reproduce qualitatively the experimental PB signal increase, the change of sign at $\hbar\omega_{probe}/2$, and the PA decrease for $E_{\rm F}>$400meV, Fig.\ref{f:gatingDT}d.

Our data and model indicate that the PA signal amplitude is maximized at $E_{\rm F}\sim0.4\hbar\omega_{probe}$, unlike what claimed in Ref.\citenum{Katayama2020}, \textit{i.e.} that the signal maximum occurs at $E_{\rm F}= 0.5\hbar\omega_{pump}$. We attribute this difference to the coarser sampling of $E_{\rm F}$ in Ref.\citenum{Katayama2020} (50meV steps against our 4meV, Fig.3b), to the fact that Ref.\citenum{Katayama2020} used a reflection geometry, mixing contributions from transient reflection and transmission with opposite sign\cite{Yao2015}, and to the larger probe photon energy at which intraband transitions contribute with opposite sign to that of interband transitions\cite{Malard2013}. Furthermore, the $E_{\rm F}$ thresholds that control the $\Delta T/T$ amplitude and sign are identified by $\hbar\omega_{probe}$, not by $\hbar\omega_{pump}$ as incorrectly stated in Ref.\citenum{Katayama2020}, since $\hbar\omega_{pump}$ has no role in defining the cooling dynamics of the HFD after the first ultrafast ($<$100fs) step of electron-electron thermalization\cite{Brida2013,Tomadin2013}.

To examine the dependence of the cooling dynamics on $E_{\rm F}$, we monitor $\Delta T/T$ as a function of pump-probe delay. Figs.\ref{f:dyn}a,c show the gate-dependent relaxation dynamics at $\hbar\omega_{probe}$=0.729, 1.033eV, lower and higher than the pump photon energy. At both energies, the relaxation dynamics progressively slows down with increasing $|E_{\rm F}|$, evolving from a biexponential to a monoexponential decay, due to a reduction of the fast decay component. We can appreciate this slowdown by noting that, to see a signal reduction by a factor 10, we need to wait$\sim$1ps at $|E_{\rm F}|=$100meV and$\sim$5ps at 300meV. Both signal intensity and relaxation dynamics are symmetric for \textit{n-} or \textit{p-}doping, as a consequence of the CB, VB symmetry.

The observed gate-dependence can be qualitatively explained considering that, for increasing $|E_{\rm F}|$, the excess energy of the photoexcited charge carriers with respect to equilibrium is reduced, affecting the scattering with optical phonons that drives the cooling. To gain a deeper insight into the phenomena responsible for quenching the fast relaxation component, we solve a set of phenomenological equations of motion (EOMs)\cite{Rana2009} for the electronic temperature, $T_{e}$, and for the occupation of the phonon modes. We include the optical phonon modes at the K and $\Gamma$ points of the SLG Brillouin zone, and we consider that they can be emitted/absorbed by e and h and decay into acoustic modes due to anharmonic coupling\cite{Lazzeri2005,Bonini2007,Tomadin2013} (see Methods).

We calculate the time-evolution of the differential conductivity for several values of the chemical potential, $\mu_c$, corresponding to $E_{\rm F}$, (i.e, $\mu_c$ at T$_e$=0\cite{Kittel}), in the range 250 to -650meV. The results in Figs.\ref{f:dyn}b,d explain the observed slowdown of the dynamics with increasing $E_{\rm F}$ with the saturation of the phase space for optical phonon-emitting electronic transitions. As $E_{\rm F}$ increases, there are fewer carriers with an energy high enough ($>$160meV) to emit an optical phonon, and optical phonon emission is quenched. This is a fundamental process, not dependent on SLG substrate, like supercollision cooling through defects\cite{Song2010}, nor on its dielectric environment, like the cooling to hyperbolic phonons in hBN-encapsulated SLG\cite{tielrooij2018}. It is determined by the intrinsic coupling of e with the K and $\Gamma$ phonons\cite{Piscanec2004}. The initial increase of PB amplitude with $E_{\rm F}$ in Figs.\ref{f:gatingDT}b,c is a consequence of the quenching of relaxation \textit{via} optical phonons\cite{Brida2013,Tomadin2013}, which reduces the initial fast decay. 

Fig.\ref{f:dyn} also shows that the Fermi energy $|E_{\rm F}^0|$ at which the $\Delta T/T$ signal changes sign is independent of $t$, both in experiments and simulations. The vanishing $\Delta T/T$ does not correspond to zero absorption, but it means that the conductivity remains at its equilibrium value for all delays. For $t>$0, the e system is photoexcited. This can happen only because the e distribution undergoes a time-evolution such that the conductivity remains time-independent at $\hbar\omega=2|E_{\rm F}|$.
\begin{figure}
\centerline{\includegraphics[]{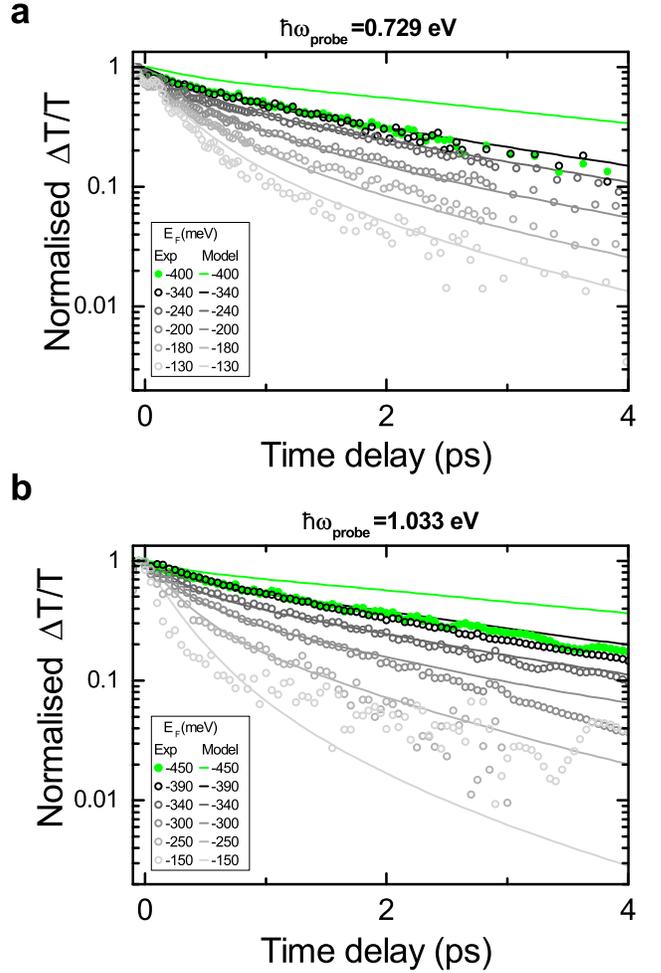}}
\caption{(\textbf{a}-\textbf{b}) Experimental (coloured dots) and theoretical (solid lines) $\Delta T/T$ at different $E_{\rm F}$  for (a) $\hbar\omega_{probe}$=0.729eV and (b) $\hbar\omega_{probe}$=1.033eV for pump-probe time delays between -500fs and 5ps (pump arrival at $t=$0).}
\label{f:dynComp}
\end{figure}
\begin{figure*}
\centerline{\includegraphics[width=160mm]{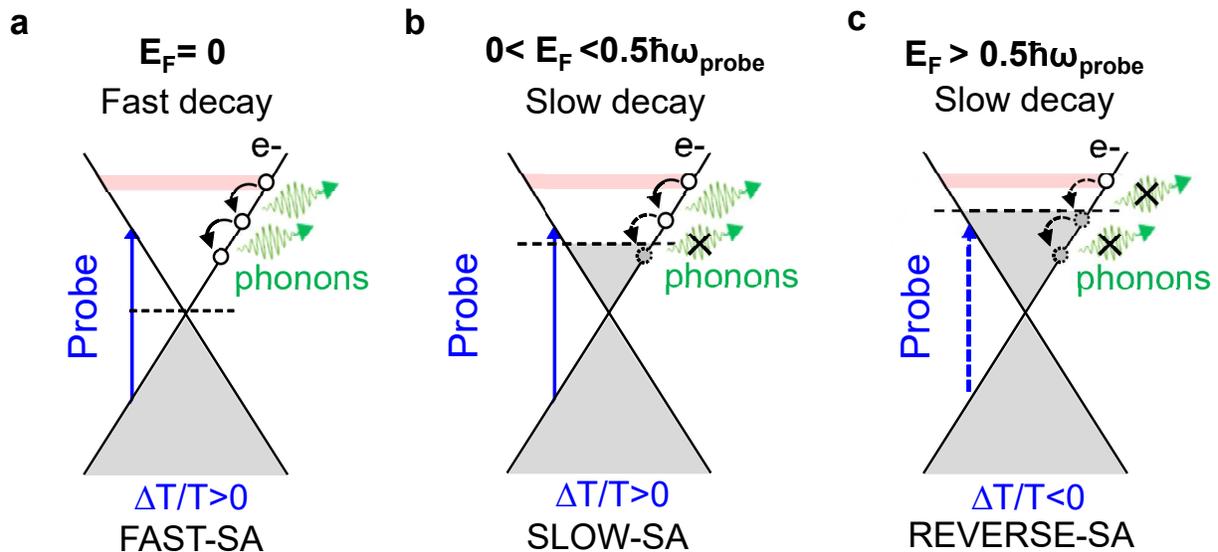}}
\caption{(\textbf{a-c}) Sketch of interband absorption of a NIR probe pulse (vertical blue arrow) within the SLG Dirac cones populated at equilibrium up to $E_{\rm F}$ (grey filling). The pump pulse perturbs the probe absorption by promoting e from VB to CB (red filling), which then relax through emission of optical phonons (downward black arrows). The three sketches correspond to (a) $E_{\rm F}$ at the Dirac point, (b,c) moderate \textit{n-}doping with $E_{\rm F}$ (b) below and (c) above the threshold for interband probe absorption. By increasing $E_{\rm F}$, optical phonon emission is quenched (dashed downward arrows) and relaxation becomes slower. Above the threshold for interband absorption of the probe (dashed vertical blue arrow), the photoexcitation results in $\Delta T/T<0$, leading to reverse saturable absorption, consisting in an increased absorption upon increasing illumination.}
\label{f:PA}
\end{figure*}

Fig.\ref{f:dynComp}a shows that for $E_{\rm F}>$340meV and $\hbar\omega_{probe}$=0.729eV, the simulations predict a further slowdown of the relaxation dynamics, not observed in our experiments. These all saturate to a similar decay trend independent on $E_{\rm F}$ (see overlapping black and green dots in Fig.\ref{f:dynComp}). Analogous behavior is found at all $\hbar\omega_{probe}$, provided that when we increase $\hbar\omega_{probe}$, we tune $E_{\rm F}$ to higher levels to find overlapping decay dynamics (see Fig.\ref{f:dynComp}b). To understand the saturation of this slowdown, we need to consider that additional relaxation channels may start playing a role once the cooling \textit{via} optical phonons gets slower. Defects can accelerate cooling\cite{graham2013photocurrent,betz2013supercollision,Song2012,alencar2014}, mediating the scattering with acoustic phonons of finite momentum and energy\cite{Song2012}. This supercollision mechanism\cite{graham2013photocurrent,betz2013supercollision,Song2012,alencar2014} may become the dominating process once optical phonon emission is quenched. Cooling times$\sim$4ps are expected for supercollision cooling\cite{graham2013photocurrent} in SLG with $E_{\rm F}\sim$400meV and mobility of a few thousand cm$^2$ V$^{-1}$ s$^{-1}$, as that of our device. The $E_{\rm F}$ independence of the decay dynamics in the high $E_{\rm F}$ limit could be explained by the lack of dependence on carrier density of the supercollision cooling time away from the Dirac point\cite{graham2013photocurrent}. According to Refs.\citenum{EEDop2009,Bruna2014}, the e scattering time with defects in SLG is not expected to significantly change with $E_{\rm F}$.

The electrical tunability of the SLG relaxation dynamics, sketched in Figs.\ref{f:PA}a-c, is promising for the realization of tunable SA. Saturable absorption, \textit{i.e.} the quenching of optical absorbance under intense illumination\cite{RP}, can occur in SLG at low light intensity (\textit{e.g.}$\sim$0.750MW cm$^{-2}$ at 0.8eV\cite{Zhang2015}). We measured a saturation intensity\cite{Sun2010} $I_{\rm S}=0.5$-$1.7$MW cm$^{-2}$ for photon energies in the range$\sim$0.5-2.5eV, comparable to semiconductor saturable absorber mirrors (SESAMs) ($P=0.01$-$0.1$MW cm$^{-2}$ at 0.944eV\cite{spuhler2005sesam}), but maintained over a much broader spectral range\cite{Sun2010}. The modulation depth, defined as the maximum change in absorption\cite{RP}, can be optically tuned exploiting cross absorption modulation\cite{sheng2014tunable}. GSAs are promising for passive mode-locking\cite{Sun2010,Popa2010,sun2010stable}, Q-switching\cite{Popa2011}, and Q-switched mode-locking\cite{xie2012graphene}.

Fig.\ref{f:PA} shows that the SLG equilibrium photoresponse can be electrically tuned, providing an additional knob for controlling its SA performance in terms of modulation depth and recovery dynamics. For $E_{\rm F}<<\hbar\omega_{probe}/2$, the intrinsic bi-exponential-like relaxation dynamics makes SLG an ideal fast SA, Fig.\ref{f:PA}a. The presence of two different time scales, in analogy with SESAMs\cite{Keller2003}, is considered an advantage for mode locking\cite{Keller2003}. As discussed in Refs.\citenum{Keller1996,Keller2003}, the longer time scale reduces the saturation intensity, facilitating self-starting mode-locking, while the fast relaxation component is efficient in shaping sub-ps pulses. For $E_{\rm F}\leq\hbar\omega_{probe}/2$ as in Fig.\ref{f:PA}b, SLG can act as slow SA\cite{Paschotta2001} with recovery time 10 to 30 times longer than the pulse duration\cite{Kartner1996,Paschotta2001}, favouring soliton shaping\cite{Kartner1996}, or the temporal shift of the pulses caused by the SA\cite{Keller2003}, which limits the time in which noise behind the pulse can be amplified\cite{Paschotta2001}. Longer recovery time also gives an increased tolerance towards instability induced by self-phase modulation\cite{Paschotta2001}.

The PA at $E_{\rm F}>\hbar\omega_{probe}/2$ can be exploited to operate SLG as reverse SA\cite{Band1986}, for which absorption increases with increasing impinging intensity, due to depletion of final state population, see Fig.\ref{f:PA}c. The PA of highly doped SLG could be exploited to realize an optical limiter\cite{Harter1986}, based on the decrease in transmittance under high-intensity or fluence illumination. An ideal optical limiter, with the functionality of protecting delicate optical elements, should strongly attenuate intense, potentially dangerous laser beams, while exhibiting high transmittance for low-intensity light. Carbon nanotubes\cite{Wang2008} and few layer graphene\cite{Wang2009} dispersions in organic solvents have been used to prepare optical limiters. However, these rely on nonlinear scattering\cite{Tutt1993}, rather than on nonlinear absorption\cite{WangBlau2009}. The nonlinear scattering of graphene dispersions\cite{Wang2009} is based on the avalanche ionization of carbon when interacting with an incident laser pulse, and subsequent bubble formation in the solvent due to the heat released by expanding microplasmas\cite{Wang2008,Wang2009}. The 10ps PA lifetime of the nonlinear absorption of highly doped SLG is 10 times shorter that the typical timescales for thermal effects and bubbling of graphene dispersions, which are of the order of 100ps\cite{Wang2008}, allowing the application to lasers with shorter pulse duration. The nonlinear absorption in SLG is not related to a specific absorption resonance, thus it covers a broad spectral range (as shown in Fig.\ref{f:gatingDT}b where for $|E_{\rm F}|$=600meV we detect PA for photon energies in the range 0.729 to 1.127eV). The SA to reverse SA transition could be used for all-optical logic gates\cite{porzi2008all}. Gate-dependent effects on cooling dynamics are also important for the design of transceivers for data communication\cite{romagnoli2018graphene}.$\textit{E.g.}$, Ref.\citenum{Castilla2019} showed that longer cooling times give larger photocurrent.
\section{Conclusions}
We demonstrated that electrostatic tuning of the non-equilibrium optical response of SLG results in changes of amplitude, sign and recovery dynamics of $\Delta T/T$. Increasing $E_{\rm F}$ quenches the emission of optical phonons, \textit{i.e.} the fastest intrinsic relaxation channel for SLG hot charge carriers. The ability to tune $E_{\rm F}$ above the threshold for Pauli blocking of interband absorption of NIR light, results in photoinduced absorption in SLG, due to pump-induced unblocking of interband transitions for the probe. Our results anticipate the use of voltage-controlled SLG for non-equilibrium optoelectronic devices, as gate tunable optical elements which can behave either as fast, slow, or reverse SA.
\section{Methods}\label{methods}
\paragraph{High-sensitivity transient absorption microscopy}
The experimental setup for the pump-probe experiments comprises a mode-locked Er-doped fiber oscillator (Toptica Photonics, FemtoFiberPro), emitting 150fs pulses at 0.8eV (1550nm) at 40MHz repetition rate. The oscillator feeds two Er-doped fiber amplifiers (EDFAs) each generating 70fs pulses at 0.8eV with 300mW average power. The output of the first EDFA is attenuated to obtain pump pulses with 1mW maximum average power. The second EDFA feeds a highly nonlinear optical fiber that produces a supercontinuum tunable between 0.729 and 1.240eV, which serves as probe pulse. The pump and probe pulses, synchronized by a computer controlled optical delay line and collinearly recombined by a dichroic beam splitter, are focused on the sample over spots of$\sim$25$\mu$m radius. The portion of the probe transmitted by the sample, spectrally selected by a monochromator with bandwidth$\sim$5nm, is detected by an amplified InGaAs photodiode (bandpass 4.5 MHz, gain 10$^4$) and analysed by a lock-in amplifier (Zurich Instruments HF). Pump and probe pulses have perpendicular polarizations and a linear polarizer is used to filter out the pump light scattered from the sample. The pump pulse is modulated at 1MHz by an acousto-optic modulator, resulting in a $\Delta T(t)/T$ sensitivity of the order of 10$^{-7}$, for an integration time of 300ms. From the FWHM of the instrumental response function we estimate an overall temporal resolution$\sim$100fs. The absorbed photon density is in the range 2-3$\times$10$^{12}$cm$^{-2}$ (depending on $E_{\rm F}$), as calculated from incident fluence and sample transmission.
	
\paragraph{Simulation of differential transmission dynamics}
To model the time-evolution of the differential transmission we assume that, on the time-scale given by the time-resolution of the experiment (100fs), e in both CB and VB are thermalized at the same $T_{\rm e}$, and reach a common chemical potential $\mu_c(t)$, such that the e energy distribution is a HFD. $\mu_c(t)$ is calculated at each instant in time, as it depends on $T_e$, and is fixed by the condition that the carrier density, defined as\cite{Kittel} $n=\int_{-\infty}^{+\infty}d\varepsilon\nu(\varepsilon) [1+exp((\varepsilon-\mu)/k_{\rm B}T)]^{-1}$, with $\nu(\varepsilon)$ the electronic density of states in SLG\cite{KatsnelsonBook}, is constant\cite{Kittel}. As for Refs.\citenum{Rana2009,Wang2010}, we can write the following EOMs for $T_{\rm e}$ and phonon occupations:
\begin{eqnarray}\label{eq:eoms}
\frac{d T_{\rm e}(t)}{d t} & = & - \frac{R_{\rm \Gamma}(t) \hbar \omega_{\rm \Gamma} + R_{K}(t) \hbar \omega_{K}}{c_{\rm e}(t) + c_{\rm h}(t)}, \nonumber \\
\frac{d n_{\Gamma}(t)}{d t} & = & \frac{R_{\Gamma}(t)}{M_{\Gamma}} - \frac{n_{\Gamma}(t) - n_{\Gamma}^{(0)}}{\tau_{\rm ph}}, \nonumber \\
\frac{d n_{K}(t)}{d t} & = & \frac{R_{K}(t)}{M_{K}} - \frac{n_{K}(t) - n_{K}^{(0)}}{\tau_{\rm ph}}.
\end{eqnarray}
Here, $n_{\Gamma}(t)$ and $n_{K}(t)$ are the occupations of the optical phonon modes at $\Gamma$ and $K$, with energy $\hbar \omega_{\Gamma}\sim0.196 {\rm eV}$ and $\hbar \omega_{K}\sim0.161{\rm eV}$\cite{Piscanec2004}, respectively, as these have the strongest electron-phonon coupling\cite{Piscanec2004}. The parameter $\tau_{\rm ph}$ is the finite optical phonon lifetime, \textit{via} relaxation into acoustic phonons due nonlinearities of the lattice\cite{Bonini2007}, until global thermal equilibrium with densities $n_{\Gamma}^{(0)}$, $n_{K}^{(0)}$ is reached. We find good agreement between theory and experiment for $\tau_{\rm ph}\simeq1.2{\rm ps}$, consistent with Ref.\citenum{Bonini2007}. The constant coefficients $M_{\Gamma}$, $M_{K}$ correspond to the number of phonon modes in an annular region between the minimum and maximum energy that can be exchanged with e\cite{Rana2009,Wang2010}. The time-dependent parameters $c_{\rm e}(t)$ and $c_{\rm h}(t)$ are the heat capacities of e in CB and of h in VB, respectively. The time-dependent parameters $R_{\Gamma}(t)$ and $R_{K}(t)$ are electronic relaxation rates per unit area, due to phonon emission and absorption, proportional to a Boltzmann scattering integral\cite{Rana2009,Wang2010}.

In line with our assumption that, on the time-scale probed by our experiments, a common $\mu_c(t)$ is established between CB and VB, the heat capacities are calculated separately in the two bands, (\textit{i.e.} $T_{\rm e}$ variations are decoupled from inter-band transitions) and only intra-band transitions are included in the relaxation rates. The initial electronic temperature $T_{\rm e}(0)$, following the pump pulse, is estimated as for Ref.\citenum{Soavi2018}. The initial phonon populations $n_{\Gamma,K}(0)$ are evaluated at RT. The optical photo-conductivity $\Delta \sigma(t)=\sigma(t)-\sigma(0)$\cite{KatsnelsonBook} depends on $T_{\rm e}(t)$ and $\mu_c(t)$. We use the Tinkham formula\cite{Tinkham1956} to obtain the differential transmission.
\section{Notes}
The authors declare no competing financial interest.
\section{acknowledgements}
We acknowledge discussions with Alessandro Principi.
We acknowledge funding from the European Union
Horizon 2020 Programme under Grant Agreement No.
881603 Graphene Core 3, ERC Grants Hetero2D and GSYNCOR, EPSRC Grants EP/K01711X/1, EP/K017144/1, EP/N010345/1, EP/L016087/1, EP/V000055/1, the German Research Foundation DFG (CRC 1375 NOA) and the Daimler und Benz foundation. K.J.T. acknowledges funding from the European Union's Horizon 2020 research and innovation program under Grant Agreement No. 804349 (ERC StG CUHL), RYC fellowship No. RYC-2017-22330, and IAE project PID2019-111673GB-I00. ICN2 was supported by the Severo Ochoa program from Spanish MINECO Grant No. SEV-2017-0706.

\end{document}